\documentclass[twocolumn,showpacs, prl]{revtex4}
\usepackage{graphicx}
\usepackage{dcolumn}
\usepackage{amsmath}
\usepackage{amssymb}

\newcommand{\SK}[0]{Super-Kamiokande-I}

\begin{document}


\title{Limits On the Neutrino Magnetic Moment Using
1496 Days of Super-Kamiokande-I Solar Neutrino Data}

\newcounter{foots}
\newcounter{notes}


\newcommand{\authoraticrr}{$^{1}$}
\newcommand{\authoratncen}{$^{2}$}
\newcommand{\authoratbu}{$^{3}$}
\newcommand{\authoratbnl}{$^{4}$}
\newcommand{\authoratuci}{$^{5}$}
\newcommand{\authoratcsu}{$^{6}$}
\newcommand{\authoratcnu}{$^{7}$}
\newcommand{\authoratgmu}{$^{8}$}
\newcommand{\authoratgifu}{$^{9}$}
\newcommand{\authoratuh}{$^{10}$}
\newcommand{\authoratkek}{$^{11}$}
\newcommand{\authoratkobe}{$^{12}$}
\newcommand{\authoratkyoto}{$^{13}$}
\newcommand{\authoratlanl}{$^{14}$}
\newcommand{\authoratlsu}{$^{15}$}
\newcommand{\authoratumd}{$^{16}$}
\newcommand{\authoratmit}{$^{17}$}
\newcommand{\authoratduluth}{$^{18}$}
\newcommand{\authoratsuny}{$^{19}$}
\newcommand{\authoratnagoya}{$^{20}$}
\newcommand{\authoratniigata}{$^{21}$}
\newcommand{\authoratosaka}{$^{22}$}
\newcommand{\authoratseoul}{$^{23}$}
\newcommand{\authoratshizuokaseika}{$^{24}$}
\newcommand{\authoratshizuoka}{$^{25}$}
\newcommand{\authoratskku}{$^{26}$}
\newcommand{\authorattohoku}{$^{27}$}
\newcommand{\authorattokyo}{$^{28}$}
\newcommand{\authorattokai}{$^{29}$}
\newcommand{\authorattit}{$^{30}$}
\newcommand{\authoratwarsaw}{$^{31}$}
\newcommand{\authoratuw}{$^{32}$}

\newcommand{\addressoficrr}[1]{$^{1}$ #1 }
\newcommand{\addressofncen}[1]{$^{2}$ #1 }
\newcommand{\addressofbu}[1]{$^{3}$ #1 }
\newcommand{\addressofbnl}[1]{$^{4}$ #1 }
\newcommand{\addressofuci}[1]{$^{5}$ #1 }
\newcommand{\addressofcsu}[1]{$^{6}$ #1 }
\newcommand{\addressofcnu}[1]{$^{7}$ #1 }
\newcommand{\addressofgmu}[1]{$^{8}$ #1 }
\newcommand{\addressofgifu}[1]{$^{9}$ #1 }
\newcommand{\addressofuh}[1]{$^{10}$ #1 }
\newcommand{\addressofkek}[1]{$^{11}$ #1 }
\newcommand{\addressofkobe}[1]{$^{12}$ #1 }
\newcommand{\addressofkyoto}[1]{$^{13}$ #1 }
\newcommand{\addressoflanl}[1]{$^{14}$ #1 }
\newcommand{\addressoflsu}[1]{$^{15}$ #1 }
\newcommand{\addressofumd}[1]{$^{16}$ #1 }
\newcommand{\addressofmit}[1]{$^{17}$ #1 }
\newcommand{\addressofduluth}[1]{$^{18}$ #1 }
\newcommand{\addressofsuny}[1]{$^{19}$ #1 }
\newcommand{\addressofnagoya}[1]{$^{20}$ #1 }
\newcommand{\addressofniigata}[1]{$^{21}$ #1 }
\newcommand{\addressofosaka}[1]{$^{22}$ #1 }
\newcommand{\addressofseoul}[1]{$^{23}$ #1 }
\newcommand{\addressofshizuokaseika}[1]{$^{24}$ #1 }
\newcommand{\addressofshizuoka}[1]{$^{25}$ #1 }
\newcommand{\addressofskku}[1]{$^{26}$ #1 }
\newcommand{\addressoftohoku}[1]{$^{27}$ #1 }
\newcommand{\addressoftokyo}[1]{$^{28}$ #1 }
\newcommand{\addressoftokai}[1]{$^{29}$ #1 }
\newcommand{\addressoftit}[1]{$^{30}$ #1 }
\newcommand{\addressofwarsaw}[1]{$^{31}$ #1 }
\newcommand{\addressofuw}[1]{$^{32}$ #1 }

\author{
{\large The Super-Kamiokande Collaboration} \\ 
\bigskip
D.W.~Liu\authoratuci,
Y.~Ashie\authoraticrr,
S.~Fukuda\authoraticrr,
Y.~Fukuda\authoraticrr,
K.~Ishihara\authoraticrr,
Y.~Itow\authoraticrr,
Y.~Koshio\authoraticrr,
A.~Minamino\authoraticrr,
M.~Miura\authoraticrr,
S.~Moriyama\authoraticrr
% ...
\\
\smallskip
\footnotesize
\it
\addressoficrr{Kamioka Observatory, Institute for Cosmic Ray Research, University of Tokyo, Kamioka, Gifu, 506-1205, Japan}\\
\addressofncen{Research Center for Cosmic Neutrinos, Institute for Cosmic Ray Research, University of Tokyo, Kashiwa, Chiba 277-8582, Japan}\\
\addressofbu{Department of Physics, Boston University, Boston, MA 02215, USA}\\
\addressofbnl{Physics Department, Brookhaven National Laboratory, Upton, NY 11973, USA}\\
\addressofuci{Department of Physics and Astronomy, University of California, Irvine, Irvine, CA 92697-4575, USA }\\
\addressofcsu{Department of Physics, California State University, Dominguez Hills, Carson, CA 90747, USA}\\
\addressofcnu{Department of Physics, Chonnam National University, Kwangju 500-757, Korea}\\
\addressofgmu{Department of Physics, George Mason University, Fairfax, VA 22030, USA }\\
\addressofgifu{Department of Physics, Gifu University, Gifu, Gifu 501-1193, Japan}\\
\addressofuh{Department of Physics and Astronomy, University of Hawaii, Honolulu, HI 96822, USA}\\
\addressofkek{High Energy Accelerator Research Organization (KEK), Tsukuba, Ibaraki 305-0801, Japan }\\
\addressofkobe{Department of Physics, Kobe University, Kobe, Hyogo 657-8501, Japan}\\
\addressofkyoto{Department of Physics, Kyoto University, Kyoto 606-8502, Japan}\\
\addressoflanl{Physics Division, P-23, Los Alamos National Laboratory, Los Alamos, NM 87544, USA }\\
\addressoflsu{Department of Physics and Astronomy, Louisiana State University, Baton Rouge, LA 70803, USA }\\
\addressofumd{Department of Physics, University of Maryland, College Park, MD 20742, USA }\\
\addressofmit{Department of Physics, Massachusetts Institute of Technology, Cambridge, MA 02139, USA}\\
\addressofduluth{Department of Physics, University of Minnesota, Duluth, MN 55812-2496, USA}\\
\addressofsuny{Department of Physics and Astronomy, State University of New York, Stony Brook, NY 11794-3800, USA}\\
\addressofnagoya{Department of Physics, Nagoya University, Nagoya, Aichi 464-8602, Japan}\\
\addressofniigata{Department of Physics, Niigata University, Niigata, Niigata 950-2181, Japan }\\
\addressofosaka{Department of Physics, Osaka University, Toyonaka, Osaka 560-0043, Japan}\\
\addressofseoul{Department of Physics, Seoul National University, Seoul 151-742, Korea}\\
\addressofshizuokaseika{International and Cultural Studies, Shizuoka Seika College, Yaizu, Shizuoka, 425-8611, Japan}\\
\addressofshizuoka{Department of Systems Engineering, Shizuoka University, Hamamatsu, Shizuoka 432-8561, Japan}\\
\addressofskku{Department of Physics, Sungkyunkwan University, Suwon 440-746, Korea}\\
\addressoftohoku{Research Center for Neutrino Science, Tohoku University, Sendai, Miyagi 980-8578, Japan}\\
\addressoftokyo{The University of Tokyo, Tokyo 113-0033, Japan }\\
\addressoftokai{Department of Physics, Tokai University, Hiratsuka, Kanagawa 259-1292, Japan}\\
\addressoftit{Department of Physics, Tokyo Institute for Technology, Meguro, Tokyo 152-8551, Japan }\\
\addressofwarsaw{Institute of Experimental Physics, Warsaw University, 00-681 Warsaw, Poland }\\
\addressofuw{Department of Physics, University of Washington, Seattle, WA 98195-1560, USA}\\
}
\affiliation{ } 

\begin{abstract}

A search for a non-zero neutrino magnetic moment has been conducted using 1496 live days
of solar neutrino data from {\SK}.  Specifically, we searched for distortions to the energy
spectrum of recoil electrons arising from magnetic scattering due to a non-zero neutrino
magnetic moment. In the absence of clear signal, we found $\mu_{\nu} \leq 3.6 \times
10^{-10}$ $\mu_{B}$ at 90\% C.L. by fitting to the Super-Kamiokande day/night spectra. The
fitting took into account the effect of neutrino oscillation on the shapes of energy spectra.
With additional information from other solar neutrino and KamLAND experiments constraining
the oscillation region, a limit of $\mu_{\nu} \leq 1.1 \times 10^{-10}$ $\mu_{B}$ at 90\% C.L.
was obtained. 

\end{abstract}

%
%



\maketitle


In the Standard Model, neutrinos are massless and do not have magnetic moments. There is now
strong evidence from recent experiments \cite{osc1,osc2,osc3,sno,osc4} that neutrinos undergo
flavor oscillations and hence must have finite masses. Introducing neutrino masses to the 
Standard Model results in neutrino magnetic moments \cite{lee}. However, such moments are at
least eight orders of magnitude below currently accessible experimental limits. These limits
are derived from reactor \(\bar{\nu}_{e}\)'s \cite{rei, der, texo, munu} and are in the range
of $(1-4) \times 10^{-10}$ $\mu_{B}$, where $\mu_{B}$ is Bohr magneton. Therefore, a positive
observation of such large magnetic moments would imply additional physics beyond the Standard
Model.

The general interaction of neutrino mass eigenstates j and k with a magnetic field can be 
characterized by constants $\mu_{jk}$, the magnetic moments. Both diagonal ($j = k$) and 
off-diagonal ($j \neq k$) moments are possible.

While there have been attempts to use the neutrino magnetic moments to explain the solar
neutrino problem \cite{vol},  {\it e.g.} spin flavor precession (SFP) \cite{lim}, SFP cannot 
explain the suppressed reactor anti-neutrino flux detected at KamLAND \cite{osc4}. Under the
assumption of CPT invariance, KamLAND's results give independent support to neutrino
oscillations \cite{osc2,osc3}, not SFP \cite{bar}, being the solution to the solar neutrino
problem.

In this paper, we report a search for neutrino magnetic moment using the high statistics solar 
neutrino data obtained by {\SK}. Super-Kamiokande (SK) is a water Cherenkov detector, with a
fiducial volume of 22.5 kton, located in the Kamioka mine in Gifu, Japan. Descriptions of the
detector can be found elsewhere \cite{nim}. SK detects solar neutrinos via the elastic
scattering of neutrinos off electrons in the water. The scattered recoil electrons are detected
via Cherenkov light, allowing their direction, timing and total energy to be measured. SK
measures the spectrum of the recoiling electrons with high statistical accuracy. To control 
energy-related systematic effects, the number of hit photomultiplier tubes (PMT) is related to
the total electron energy using electrons injected by an electron linear accelerator (LINAC)
\cite{linac}. The number of hit PMTs in the Monte Carlo simulation of those LINAC electrons is
tuned to agree with LINAC data. As a result of this tuning, the systematic uncertainty of the
reconstructed energy of electrons between 5 and 20 MeV is less than 0.64\%. The uncertainty of
the energy resolution is less than 2\%. This absolute energy scale is monitored and cross
checked by (1) muon decay electrons, (2) spallation products induced by cosmic ray muons, and
(3) decay of artificially produced $^{16}\mbox{N}$ \cite{dt}. The data used for this analysis
were collected from May 31, 1996 to July 15, 2001 with a livetime of 1496 days. The results
are binned in 0.5 MeV bins of the total electron energy from 5 to 14 MeV and one bin combining
events from 14 to 20 MeV. As a real time detector, SK can divide the data sample into day and
night data samples which give the day/night spectra. The number of events in each energy bin is
extracted individually by utilizing the directional correlation between the recoil electrons 
and the Sun. The angular distribution in the region far from the solar direction is used to
estimate the background. The estimation of the backgrounds, along with the expected angular
distributions of the solar neutrino signals, are incorporated into an extended maximum
likelihood method to extract the number of solar neutrino events \cite{osc2}.  

If $\mu_{\nu} \neq 0$, the differential cross section of neutrino-electron scattering is an
incoherent sum of weak scattering (\ref{weak}) and magnetic scattering (\ref{magnetic})
\cite{bea}.

\begin{equation}
\label{weak}
\left( \frac{d\sigma}{dT} \right)_{\mbox{\small{\it WK}}} = \: C  \left[
g_{L}^{2} + g_{R}^{2} \: (\: 1 - \frac{T}{E_{\nu}} \: )^2 - g_{L}g_{R}
\frac{m_{e}T}{E_{\nu}^{2}} \right]
\end{equation}

where 
$C = 2G_{F}^{2}m_{e}/\pi$, 
\(g_{L} = \sin^{2}\theta_{W} + 1/2\) for \(\nu_{e}\), 
\(g_{L} = \sin^{2}\theta_{W} - 1/2\) for \(\nu_{\mu}\) and \(\nu_{\tau}\), 
and \(g_{R} = \sin^{2}\theta_{W}\).

\begin{equation}
\label{magnetic}
\left(\:\frac {d\sigma}{dT}\:\right)_{\mbox{\small{\it EM}}} = \mu_{\nu}^{2}\: 
\frac{\pi \alpha^{2}_{em}}{m_{e}^{2}}\; 
\left( \; \frac {1}{T} - \frac {1}{E_{\nu}} \: \right)
\end{equation}

where \(\mu_{\nu}\) is in units of \(\mu_{B}\), $E_{\nu}$ is the neutrino energy, 
$T = E_{e} - m_{e}$, and $T$($E_{e}$) is the kinetic (total) energy of the recoil electrons.

\begin{figure}[t]
\center{
\includegraphics[width=8.5cm,clip]{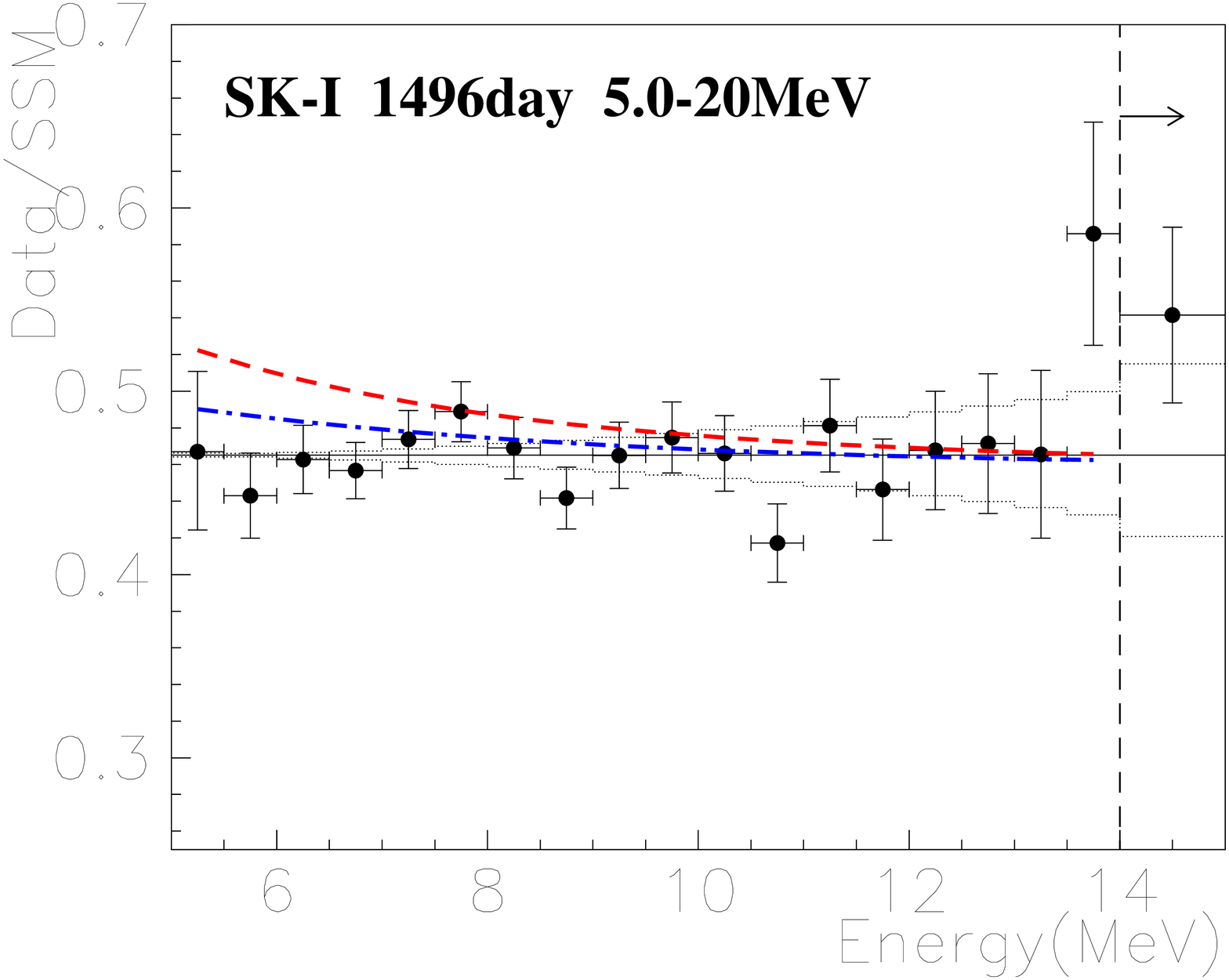}}
\caption{Ratio of SK-I observed recoil electron energy spectrum and the expected non-oscillated
weak scattering spectrum. The error bars are the results of the statistical and energy
non-correlated systematic errors being added in quadrature. The dotted lines are the
correlated systematic errors. The dash-dotted line is the expected oscillated weak scattering
spectrum for $\Delta m^{2} =6.6 \times 10^{-5} \mbox{eV}^{2}, \tan^{2}\theta = 0.48$. The
dashed line shows the addition of magnetic scattering with $\mu_{\nu} = 1.1 \times 10^{-10}
\mu_{B}$ on top of the oscillated weak spectrum. (The zero has been suppressed).}
\label{spectrum}
\end{figure}

We search for the effects of the neutrino magnetic moments by looking for distortions in the
shape of the recoil electron spectrum relative to the expected weak scattering spectrum.
Figure \ref{spectrum} shows the ratio of SK measured recoil electron energy spectrum and 
the expected weak scattering spectrum assuming no oscillation. It is flat, with no obvious 
increase of event rates in the lower energy bins. As neutrino oscillation could change the
expected weak scattering spectrum, the flatness could be due to a combination of a decrease of
the weak scattering rate by oscillation and an increase of the magnetic scattering rate at
lower energies. To investigate this the observed SK day/night energy spectra are examined using
the following \(\chi^2\), similar to the one used in SK's standard solar spectrum analysis
\cite{fuk} with the addition of the oscillation effects and the contribution of magnetic
scattering:

\begin{equation}
\chi^{2} = \sum_{a=d,n} \sum_{i=1}^{18}\; \left[ \; \frac { \frac {\alpha}{1 + 
\beta \: \delta_{i}} \: \left( W_{i}^{a} + \mu_{10}^{2} M_{i} \right) - 
D_{i}^{a}} {\sigma_{i}^{a}} \; \right] ^{2} + \beta^{2} 
\end{equation}

where  $W_{i}^{d,n}$  is the ratio of the oscillated day/night weak scattering spectra to the 
non-oscillated weak scattering spectrum. We approximate the solar neutrino oscillations by a
two-neutrino description with parameters $\theta$ (mixing angle) and $\Delta m^{2}$
(difference in mass squared) \cite{osc2}. $M_{i}$ is the ratio of the magnetic scattering
spectrum to the non-oscillated weak scattering spectrum assuming $\mu_{\nu} = 10^{-10}
\mu_{B}$. $D_{i}^{d,n}$ is the ratio of the measured day/night spectra to the non-oscillated
weak scattering spectrum.  $\delta_{i}$ is the energy bin correlated systematic error,
$\sigma_{i}^{d,n}$ is the day/night statistical and uncorrelated systematic errors added 
quadratically. $\alpha$ is the normalization factor of the measured $^{8}$B flux to the
expected flux; the $^{8}$B flux is not constrained in this analysis. $\mu_{10}^{2}$ is magnetic
moment squared in units of $(10^{-10}$ $\mu_{B})^{2}$. $\beta$ is a parameter used to
constrain the variation of correlated systematic errors which come from the uncertainties in
the energy scale, resolution and $^8$B neutrino energy spectrum. Considering neutrinos with
only diagonal magnetic moments \cite{bea}, the survival probability of neutrinos passing
through the magnetic field in the Sun is independent of neutrino energy \cite{moh}. Thus the
shape of the $^{8}$B neutrino spectrum will not be changed by the magnetic field in the Sun.
In this paper we use the SK day/night spectra from 5 to 14 MeV and consider only the $^8$B
solar neutrino flux. Furthermore, we assume $\mu_{\nu1} = \mu_{\nu2}$, so the magnetic
scattering spectrum would not be affected by neutrino oscillations. 

The $\chi^{2}$ is minimized with respect to the parameters $\alpha$, $\beta$ and
$\mu_{\nu}^{2}$ in the whole oscillation parameter space. We impose the physical condition
$\mu_{\nu}^{2} \ge 0$ in the process of minimization. As there is no strong distortion of the
observed energy spectra, this $\chi^{2}$ can be used to exclude certain regions in the
oscillation parameter space. 

\begin{figure}[ht]
\center{
\includegraphics[width=9.9cm,clip]{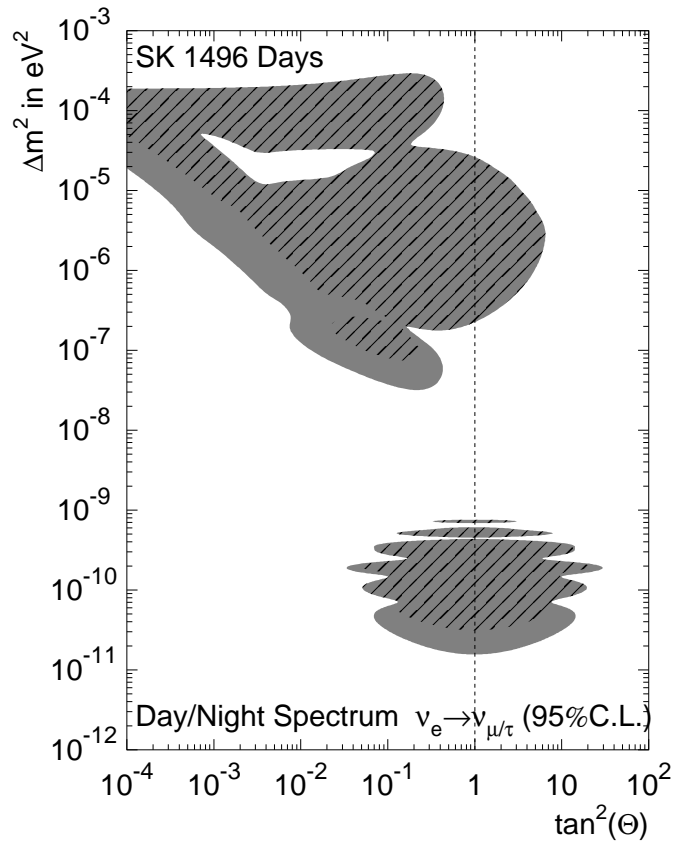}}
\caption{$95\%$ C.L. exclusion regions using the SK day/night spectra shape. The shaded area
assumes only weak scattering. The hatched region takes into account the contribution from
magnetic scattering.}
\label{exclusion}
\end{figure}

\begin{figure}[t]
\center{
\includegraphics[width=7.5cm,clip]{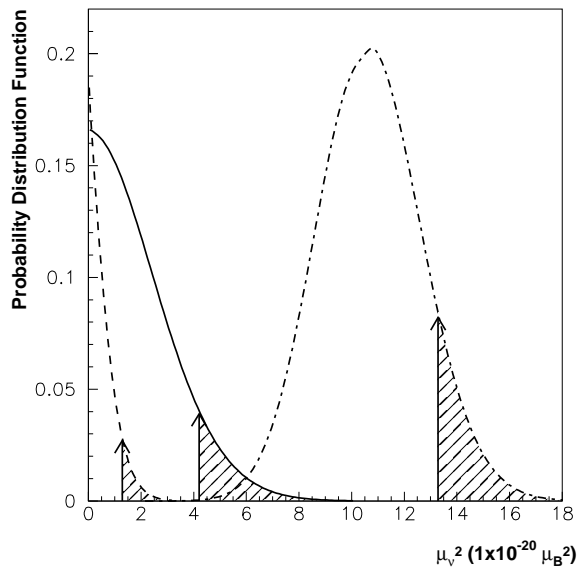}}
\caption{The probability distribution of $\Delta\chi^{2}$ as function of $\mu_{\nu}^{2}$. The 
solid line is for the case of no oscillation. The dashed line is for $\Delta m^{2} = 2.8 \times
10^{-5} \mbox{eV}^{2}$ and $\tan^{2} \theta = 0.42$ (LMA). The dash-dotted line is for $\Delta
m^{2} = 3.13 \times 10^{-11} \mbox{eV}^{2}$ and $\tan^{2} \theta = 0.91$ (VAC). The arrows
point to the place where the 90\% C.L. limits are. The hatched areas to the right of the arrows
are 10\% of the total areas under the curves.}
\label{chi2}
\end{figure}

In Figure \ref{exclusion}, the shaded regions are excluded by SK day/night spectra at $95 \%$
C.L. considering only weak scattering, while the hatched regions are excluded at the same
confidence level but including the contribution from the magnetic scattering. The exclusion
regions shrink with the addition of the magnetic scattering because there is one more
parameter with which to minimize the $\chi^{2}$. As there is no obvious increase of event rates
at lower energies, we instead derive a limit on the neutrino magnetic moment. For each point
in the oscillation parameter space, the probability distribution of $\Delta \chi^{2} = \chi^{2}
 - \chi^{2}_{min}$ as a function of the square of the magnetic moment is used. Figure 
\ref{chi2} shows the probability distributions of $\Delta \chi^{2} $ as function of 
$\mu_{\nu}^{2}$ for some oscillation parameters. 

A 90\% C.L. upper limit $\mu_{0}$ on the neutrino magnetic moment is obtained by Equation
\ref{prob} for each point in the oscillation parameter space. 

\begin{equation}
\label{prob}
\mbox{Prob}(\Delta \chi^{2}(\mu^{2} \ge \mu^{2}_{0})) = 
0.1 \times \mbox{Prob}(\Delta \chi^{2}(\mu^{2} \ge 0))
\end{equation}

The overall limit on the neutrino magnetic moment is obtained by finding the maximum of the 
aforementioned limits in the oscillation parameter space. Discarding the regions excluded by 
SK day/night spectra, we found at 90\% C.L. $\mu_{\nu} \le 3.6 \times 10^{-10} \mu_{B}$ with
the limit at $\Delta m^{2} = 3.13 \times 10^{-11} \mbox{eV}^{2}$ and $\tan^{2}\theta = 0.91$
which is in the VAC region.

\begin{figure}[t]
\center{
\includegraphics[width=8cm,clip]{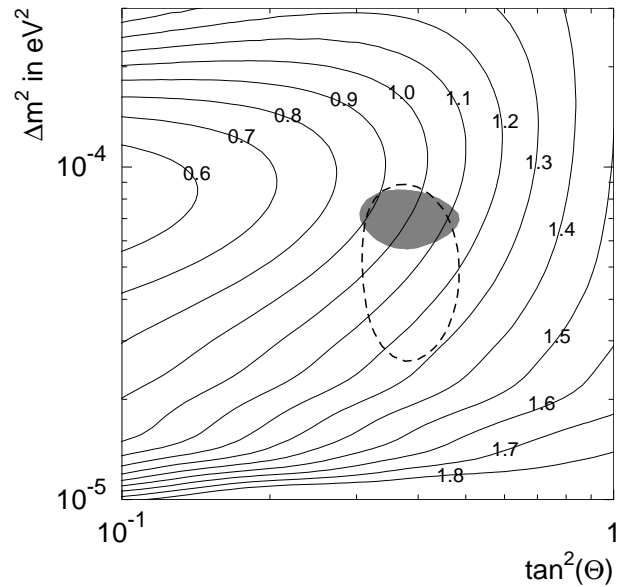}}
\caption{The 90\% C.L. $\mu_{\nu}$ limit contours (in units of $10^{-10} \mu_{B}$) and neutrino
oscillation allowed regions. The area within the dashed lines is the solar neutrino
experiments' allowed region considering both weak and magnetic scattering. The shaded area
shows the allowed region for solar experiments plus KamLAND.} 
\label{contour}
\end{figure}

Results from other solar neutrino experiments can further constrain the allowed regions in the 
oscillation parameter space. Radiochemical experiments Homestake \cite{chlorine}, SAGE
\cite{sage} and Gallex/GNO \cite{gallium} (combined into a single ``Gallium'' rate) detect
solar neutrinos via charged current interactions with nucleons. The presence of a non-zero
neutrino magnetic moment would not affect their measurements of solar neutrino flux rates. SNO
\cite{sno} extracts the charged current, neutral current and elastic scattering rates by
utilizing their distinctive angular distributions. Inclusion of the neutrino magnetic moment
will not affect the charged current interaction. The effects of non-zero neutrino magnetic
moment on the SNO neutral current interaction are estimated to be very small \cite{tsuji}.
Such a magnetic moment could change the elastic scattering rates but would not change the
angular distribution of the elastic scattering events. Therefore, SNO's charged current and
neutral current rates will be essentially unaffected by a non-zero neutrino magnetic moment.
The combination of these charged current rates with SNO's neutral current rate and SK's
day/night spectra constrains the neutrino oscillation to an area in the large mixing angle
(LMA) region as shown in Figure \ref{contour} (the area within the dashed lines).

Limiting the search for the neutrino magnetic moment within the region allowed by solar 
neutrino experiments, we get an upper limit on the neutrino magnetic moment of $\mu_{\nu} \leq 
1.3 \times 10^{-10}$ $\mu_{B}$ at 90\% C.L. with the limit at $\Delta m^{2} = 2.8 \times
10^{-5} \mbox{eV}^{2}, \tan^{2}\theta = 0.42$.

KamLAND uses inverse $\beta$-decay interactions to detect reactor \(\bar{\nu}_{e}\)'s 
\cite{osc4}. The signature of magnetic scattering with non-zero neutrino magnetic moment bears 
no similarity to that used to detect the inverse $\beta$-decay interactions. Therefore, 
KamLAND's detection of anti-neutrinos would not be affected by a non-zero neutrino magnetic
moment. Assuming CPT invariance, the inclusion of the KamLAND results further constrains the
neutrino oscillation solutions in the LMA region (the shaded area in Figure \ref{contour}).
This results in a limit on the neutrino magnetic moment at 90\% C.L. of  $\mu_{\nu} \leq 1.1
\times 10^{-10}$ $\mu_{B}$ with the limit at $\Delta m ^{2} = 6.6 \times 10^{-5} 
\mbox{eV}^{2}$ and $\tan^{2}\theta = 0.48$. This result is comparable to the most recent
magnetic moment limits from reactor neutrino experiments of ${1.3 \times 10^{-10}}$ $\mu_{B}$
(TEXONO) \cite{texo} and ${1.0 \times  10^{-10}}$ $\mu_{B}$ (MUNU) \cite{munu}, albeit for
neutrinos and not antineutrinos.

If neutrinos have off-diagonal moments, the magnetic field in the Sun can affect the $^{8}$B 
neutrino flux spectrum, so the results on the limits of neutrino magnetic moment could in 
principle be changed. But for the LMA region, the effect of the solar magnetic field is
negligible \cite{bar,grim}, so the same limits on the neutrino magnetic moment in the LMA
region would be obtained.

{\it Conclusion.} -- Limits on the neutrino magnetic moment have been obtained by analyzing
the SK day/night energy spectra. The oscillation effects on the shape of the weak scattering 
spectrum have been taken into account when analyzing energy spectra. A limit of $3.6 \times
10^{-10}$ $\mu_{B}$ using {\SK}'s 1496 days of solar neutrino data is obtained. By constraining
the search to only the regions allowed by all neutrino experiments, a limit of $1.1 \times
10^{-10}$ $\mu_{B}$ is obtained. 

\begin{acknowledgments}

The authors gratefully acknowledge the cooperation of the Kamioka Mining and Smelting Company.
Super-Kamiokande has been built and operated from funding by the Japanese Ministry of
Education, Culture, Sports, Science and Technology, the U.S. Department of Energy, and the U.S.
National Science Foundation. This work was partially supported by the Korean Research
Foundation (BK21) and the Korea Ministry of Science and Technology.

\end{acknowledgments}

\end{document}